 \documentstyle[aps,prd,epsfig,preprint]{revtex} 
 \tightenlines
\newlength{\pushupfigure}
\def \epsfin_v1#1#2{
        \vspace{\pushupfigure}
        \center
        \leavevmode
        \epsfxsize=#1
        \epsffile[20 143 575.75 698.75]{#2}
}
\newcommand{\pt}{\mbox{$p_{T}$}}
\newcommand{\et}{\mbox{$E_{T}$}}
\newcommand{\met}{\mbox{${E\!\!\!\!/_{T}}$}}
\newcommand{\metvec}{\mbox{${\vec{E}\!\!\!\!/_{T}}$}}
\newcommand{\ppbar}{\mbox{$p\overline{p}$}}

\newcommand{ \gluino}   {\mbox{$\tilde{g}$}}
\newcommand{ \squark}   {\mbox{$\tilde{q}$}}
\newcommand{ \squarkb}   {\mbox{$\bar{\tilde{q}}$}}

\newcommand{ \chizero}	{\mbox{$\tilde{\chi}_{1}^0$}}
\newcommand{ \chione }	{\mbox{$\tilde{\chi}_{1}^{\pm}$}}
\newcommand{ \chitwo }	{\mbox{$\tilde{\chi}_{2}^{0}$}}
\newcommand{ \stau}   	{\mbox{$\tilde{\tau}$}}
\newcommand{ \sleptonR}   {\mbox{$\tilde{\ell}_{R}$}}
\newcommand{ \mgluino}  {\mbox{$M_{\gluino}$}}
\newcommand{ \msquark}  {\mbox{$M_{\squark}$}}

\newcommand{\gevc} {\mbox{${\rm GeV}/c$}}
\newcommand{\gevcc}{\mbox{${\rm GeV}/c^2$}}
\newcommand{\tevcc}{\mbox{${\rm TeV}/c^2$}}
\newcommand{\ipb}{\mbox{${\rm pb}^{-1}$}}

\newcommand{\dzero}{\mbox{${\rm D\O}$}}
\def\oversim#1#2{\lower4pt\vbox{\baselineskip0pt \lineskip1.5pt
            \ialign{$\mathsurround=0pt#1\hfil##\hfil$\crcr#2\crcr\sim\crcr}}}
\def \gtsim    {\relax\ifmmode{\mathrel{\mathpalette\oversim >}}
                  \else{$\mathrel{\mathpalette\oversim >}$}\fi}
\def \ltsim    {\relax\ifmmode{\mathrel{\mathpalette\oversim <}}
                  \else{$\mathrel{\mathpalette\oversim <}$}\fi}

\newcommand{\ISAJET}{{\sc isajet}}
\newcommand{\etal}{{\it et al.}}
\begin{document}
\draft

\title{
\boldmath Search for Gluinos and Squarks Using Like-Sign Dileptons
in \ppbar\ Collisions at $\sqrt{s} = 1.8$ TeV}
%
%
\maketitle
%
%
%
%
\font\eightit=cmti8
\def\r#1{\ignorespaces $^{#1}$}
\hfilneg
\begin{sloppypar}
\noindent
T.~Affolder,\r {23} H.~Akimoto,\r {45}
A.~Akopian,\r {37} M.~G.~Albrow,\r {11} P.~Amaral,\r 8  
D.~Amidei,\r {25} K.~Anikeev,\r {24} J.~Antos,\r 1 
G.~Apollinari,\r {11} T.~Arisawa,\r {45}
% ...
\end{sloppypar}

 \vskip .026in	
\begin{center}
(CDF Collaboration)
\end{center}
 \vskip .026in		
\begin{center}
\r 1  {\eightit Institute of Physics, Academia Sinica, Taipei, Taiwan 11529, 
Republic of China} \\
\r 2  {\eightit Argonne National Laboratory, Argonne, Illinois 60439} \\
\r 3  {\eightit Istituto Nazionale di Fisica Nucleare, University of Bologna,
I-40127 Bologna, Italy} \\
\r 4  {\eightit Brandeis University, Waltham, Massachusetts 02254} \\
\r 5  {\eightit University of California at Davis, Davis, California  95616} \\
\r 6  {\eightit University of California at Los Angeles, Los 
Angeles, California  90024} \\  
\r 7  {\eightit Instituto de Fisica de Cantabria, CSIC-University of Cantabria, 
39005 Santander, Spain} \\
\r 8  {\eightit Enrico Fermi Institute, University of Chicago, Chicago, 
Illinois 60637} \\
\r 9  {\eightit Joint Institute for Nuclear Research, RU-141980 Dubna, Russia}
\\
\r {10} {\eightit Duke University, Durham, North Carolina  27708} \\
\r {11} {\eightit Fermi National Accelerator Laboratory, Batavia, Illinois 
60510} \\
\r {12} {\eightit University of Florida, Gainesville, Florida  32611} \\
\r {13} {\eightit Laboratori Nazionali di Frascati, Istituto Nazionale di Fisica
               Nucleare, I-00044 Frascati, Italy} \\
\r {14} {\eightit University of Geneva, CH-1211 Geneva 4, Switzerland} \\
\r {15} {\eightit Glasgow University, Glasgow G12 8QQ, United Kingdom}\\
\r {16} {\eightit Harvard University, Cambridge, Massachusetts 02138} \\
\r {17} {\eightit Hiroshima University, Higashi-Hiroshima 724, Japan} \\
\r {18} {\eightit University of Illinois, Urbana, Illinois 61801} \\
\r {19} {\eightit The Johns Hopkins University, Baltimore, Maryland 21218} \\
\r {20} {\eightit Institut f\"{u}r Experimentelle Kernphysik, 
Universit\"{a}t Karlsruhe, 76128 Karlsruhe, Germany} \\
\r {21} {\eightit Center for High Energy Physics: Kyungpook National
University, Taegu 702-701; Seoul National University, Seoul 151-742; and
SungKyunKwan University, Suwon 440-746; Korea} \\
\r {22} {\eightit High Energy Accelerator Research Organization (KEK), Tsukuba, 
Ibaraki 305, Japan} \\
\r {23} {\eightit Ernest Orlando Lawrence Berkeley National Laboratory, 
Berkeley, California 94720} \\
\r {24} {\eightit Massachusetts Institute of Technology, Cambridge,
Massachusetts  02139} \\   
\r {25} {\eightit University of Michigan, Ann Arbor, Michigan 48109} \\
\r {26} {\eightit Michigan State University, East Lansing, Michigan  48824} \\
\r {27} {\eightit University of New Mexico, Albuquerque, New Mexico 87131} \\
\r {28} {\eightit The Ohio State University, Columbus, Ohio  43210} \\
\r {29} {\eightit Osaka City University, Osaka 588, Japan} \\
\r {30} {\eightit University of Oxford, Oxford OX1 3RH, United Kingdom} \\
\r {31} {\eightit Universita di Padova, Istituto Nazionale di Fisica 
          Nucleare, Sezione di Padova, I-35131 Padova, Italy} \\
\r {32} {\eightit University of Pennsylvania, Philadelphia, 
        Pennsylvania 19104} \\   
\r {33} {\eightit Istituto Nazionale di Fisica Nucleare, University and Scuola
               Normale Superiore of Pisa, I-56100 Pisa, Italy} \\
\r {34} {\eightit University of Pittsburgh, Pittsburgh, Pennsylvania 15260} \\
\r {35} {\eightit Purdue University, West Lafayette, Indiana 47907} \\
\r {36} {\eightit University of Rochester, Rochester, New York 14627} \\
\r {37} {\eightit Rockefeller University, New York, New York 10021} \\
\r {38} {\eightit Rutgers University, Piscataway, New Jersey 08855} \\
\r {39} {\eightit Texas A\&M University, College Station, Texas 77843} \\
\r {40} {\eightit Texas Tech University, Lubbock, Texas 79409} \\
\r {41} {\eightit Institute of Particle Physics, University of Toronto, Toronto
M5S 1A7, Canada} \\
\r {42} {\eightit Istituto Nazionale di Fisica Nucleare, University of Trieste/
Udine, Italy} \\
\r {43} {\eightit University of Tsukuba, Tsukuba, Ibaraki 305, Japan} \\
\r {44} {\eightit Tufts University, Medford, Massachusetts 02155} \\
\r {45} {\eightit Waseda University, Tokyo 169, Japan} \\
\r {46} {\eightit University of Wisconsin, Madison, Wisconsin 53706} \\
\r {47} {\eightit Yale University, New Haven, Connecticut 06520} \\
\r {(\ast)} {\eightit Now at Carnegie Mellon University, Pittsburgh,
Pennsylvania  15213} \\
\r {(\ast\ast)} {\eightit Now at Northwestern University, Evanston, Illinois 
60208}
\end{center}


\begin{abstract}
\par\noindent We present results of the first 
search for like-sign dilepton 
($e^{\pm}e^{\pm}$, $e^{\pm}\mu^{\pm}$, $\mu^{\pm}\mu^{\pm}$) events
associated with multijets and large missing energy using 106 \ipb\
of data in \ppbar\ collisions at $\sqrt{s} = 1.8$ TeV collected during
1992-95 by the CDF experiment.  Finding no events that pass our
selection, we examine pair-production of gluinos ($\gluino$) and
squarks ($\squark$) in a constrained framework of 
the Minimal Supersymmetric Standard Model. 
At $\tan\beta = 2$ and $\mu=-800$~\gevcc,
we set  95\% confidence level  limits of
$\mgluino > 221$~\gevcc\ for $\mgluino \simeq \msquark$, and
$\mgluino > 168$ \gevcc~for $\msquark \gg \mgluino$,
both with small variation as a function of $\mu$.
\end{abstract}

 \pacs{ }   



The Standard Model (SM) of particle physics is enormously successful in
explaining a wide variety of phenomena.
In spite of this, there are a number of structural defects in the model,
such as the quadratic mass divergence of the Higgs boson.
Supersymmetry (SUSY)  provides a promising solution and
in the  Minimal Supersymmetric Standard Model (MSSM)~\cite{mssm}
each SM particle has a SUSY  partner which is required
to be lighter than or of the order of 1~\tevcc \cite{mssm}.
Conservation of $R$-parity~\cite{rparity} requires
SUSY particles to be produced in pairs and 
 the lightest SUSY particle (LSP) to be stable.

At the Fermilab Tevatron,
pair-production and sequential decays of
supersymmetric quarks (squarks, \squark) 
and supersymmetric gluons (gluinos, \gluino)
can result in events with  final state leptons.
The $\squark$ can  decay to  the lightest chargino (\chione) or 
the next-to-lightest neutralino (\chitwo)
via  $\squark \to q^{\prime}\chione$ or $\squark \to q\chitwo$, and
the $\squark \to q\gluino$ decay occurs when kinematically allowed. 
The decays of the $\gluino$ are $\gluino \to q\bar{q^{\prime}} \chione$
or $\gluino \to q \bar{q}\chitwo$.
Each  $\squark$ and $\gluino$ decay  can eventually 
produce isolated leptons and missing transverse energy ($\met$)~\cite{cdfdefs}
via the decays 
$\chione \to \ell^{\pm} \nu_{\ell} \chizero$ or 
$\chitwo \to \ell^{\pm} \ell^{\mp} \chizero$
where  \chizero\ is the LSP \cite{ellis}
which exits the detector  without interacting.
Thus, \gluino\gluino, \gluino\squark\ and \squark$\bar{\squark}$ 
production can lead to the like-sign (LS) dilepton signatures of 
$e^{\pm}e^{\pm}$, $e^{\pm}\mu^{\pm}$ and $\mu^{\pm}\mu^{\pm}$~\cite{LSD}
with two or more jets and appreciable $\met$.
The fraction of dilepton events which are LS can be as large as 30\% in some
regions of MSSM parameter space. 

The $\ell^{\pm}\ell^{\pm} + \geq 2$ jets $ + \met$ channel
is a clean signature to search for SUSY.
It has an advantage over the opposite-sign (OS) dilepton channel as there are 
only small SM backgrounds.  
Even  without the \met\ requirement
the LS analysis  is also useful for testing other theories beyond the
SM, including $R$-parity violating SUSY \cite{rpv_lsd}.
The dilepton decay channels are a natural complement 
to other direct searches for squarks and gluinos
in the \met\  plus 
multijet channel~\cite{UA1_UA2,L3,ALEPH,CDF_MET_0_IA,D0_MET_IA,CDF_Maria}. 

In this Letter, we present the results of the first search for 
$\ell^{\pm}\ell^{\pm} + \geq 2$ jets $ + \met$ events using
106~\ipb\ of data from \ppbar\ collisions at $\sqrt{s}$ = 1.8~TeV.
The data were collected by the Collider Detector at Fermilab
(CDF)~\cite{det} during the 1992-95 run of the
Tevatron.  We briefly describe the detector subsystems relevant to this
analysis.  The location of the \ppbar\ collision event vertex
($z_{vertex}$) is measured along the beam direction with a time
projection chamber.  The \pt\ of charged
particles are measured in the region $|\eta|<1.1$ by a central
tracking chamber (CTC) which is located in a 1.4~T solenoidal
magnetic field. The momentum resolution is  
$\delta \pt / \pt^{2} \simeq 0.001$ 
where \pt\ is measured in~\gevc. 
 Electromagnetic and hadronic calorimeters are segmented in a
projective tower geometry surrounding the solenoid and 
cover the region $|\eta|<4.2$. 
A muon detector is
located outside the hadron calorimeter and covers the region
$|\eta|<1.0$. 

The analysis begins with a sample of 515,699 loosely selected dilepton 
events~\cite{thesis,TRIL1A_1B} 
from which we select an initial dilepton plus dijet sample.
To ensure that the trigger is fully efficient, 
we require each event to have a lepton with $\pt \ge 11$~\gevc\ 
and  $|\eta|<1.0$ for electrons or $|\eta|<0.6$ for muons.  
A second electron or muon is required with 
$\pt \ge 5$~\gevc\ and $|\eta|<1.0$. 
Each lepton is required to be isolated such that there is 
no more than 4~GeV of transverse energy 
(measured by the calorimeter or CTC) in a cone of 
$\Delta R \equiv \sqrt{(\Delta \eta)^2 + (\Delta \phi)^2} = 0.4$
 around the direction of the lepton. 
To ensure that both leptons originated from
the same collision event and are well measured, we require 
$|z_{vertex}| \le 60$~cm and
$|z_{lepton} - z_{vertex}| \le 5$~cm for each lepton, where
$z_{lepton}$ is measured along the beamline.
In addition to the leptons, we require two or more jets with
$\et \ge 15$~GeV and $|\eta|< 2.4$.

Since the OS sample is used as a check of our understanding 
of the LS backgrounds,
we place the same cuts on both samples in parallel,
but with additional cuts on the OS events
so as to remove events which might give a kinematic bias.
To reduce the large $J/\psi$ and $\Upsilon$ component of the background
we remove the events with $M_{\ell\ell} < 12$~\gevcc.
A total of  239 OS and 16 LS dilepton events pass the requirement.

The dominant SM backgrounds are from Drell-Yan
($\gamma^{\ast}$/$Z^0$), $t\bar{t}$, $b\bar{b}$, $c\bar{c}$, and 
diboson ($W^+W^-$, $W^{\pm}Z^0$, $Z^0Z^0$) production. 
Each is  estimated using
the \ISAJET\  Monte Carlo (MC) event generator~\cite{isajet} 
and a simulation of the CDF detector. 
The cross sections for
$\gamma^{\ast}$/$Z^0$ and $t\bar{t}$ production, 
and contributions due to $B^0\bar{B}^0$ mixing events are
normalized to CDF measurements \cite{CDF_DY_Z,CDF_TOP,bbar_mixing}. 
We use next-to-leading order (NLO)
cross-sections for diboson production~\cite{diboson}.  The
contribution from 
$W(\to \ell \nu_{\ell}) + \ge 3$~jets events where one of the jets is
misidentified as a lepton is found to be negligible.

\begin{figure}
\begin{center}
\vspace*{-0.5cm}
\epsfig{file=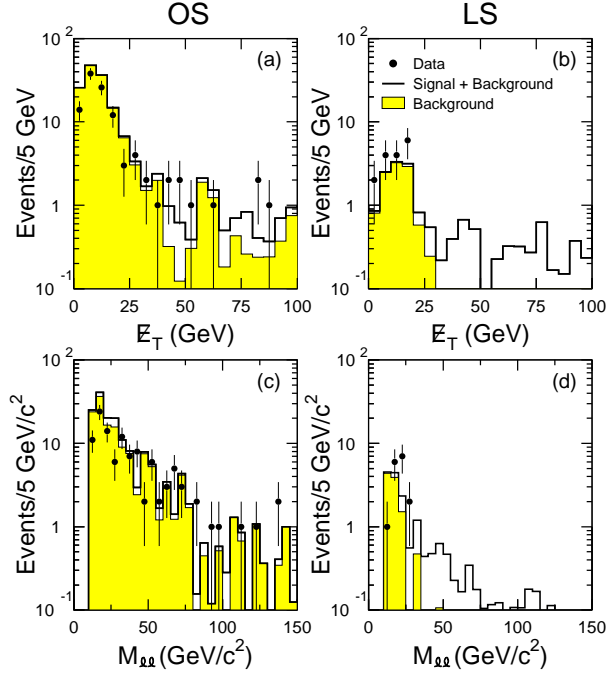,width=250pt} 
\caption{Distributions for the dilepton + dijet data  
after the $M_{\ell\ell} > 12\ \gevcc$ and  $Z^{0}$ veto requirements.
Figures (a) and (b) show the $\met$ distributions
for OS and LS samples, respectively.
The data (points) are compared to
the Standard Model background (shaded) with
a SUSY contribution  (solid) for
$\tan\beta = 2$, $\mu = -800\ \gevcc$,
$\mgluino\ = 210\ \gevcc$, and $\msquark\ = 211\ \gevcc$. 
Figures (c) and (d) show
 the $M_{\ell\ell}$ distributions in the
OS and LS samples for the same requirements.}
\label{fig:data}
\end{center}
\end{figure}

Given the large $\met$ signature from SUSY, we require at least
25~GeV of $\met$ for all dilepton events.
In the OS sample, 
we also remove all same-flavor OS dilepton events with 
$76 < M_{\ell^+\ell^-} <106$~\gevcc.
Figure~\ref{fig:data} compares the \met\ and $M_{\ell\ell}$
distributions for the data and 
the SM backgrounds for the OS and LS samples after the $Z^0$ veto but before
the \met\ requirement.
After all cuts, we observe 19 OS (4 $ee$, 10 $e\mu$, 5 $\mu\mu$) events
and no LS events
in agreement with
the SM expectation of 14.1 $\pm$ 1.3 (stat) $\pm$ 2.8 (sys) OS events 
and  0.55 $\pm$ 0.25 $\pm$ 0.08 LS events.
Tables~\ref{tab:datared} and \ref{tab:SM_BG} show a comparison of
the data reduction and
the SM backgrounds. 
There is no evidence for new particle production.

We examine the exclusion region of \msquark\ and \mgluino\
in a constrained framework of the MSSM.
We assume five squarks
($\tilde{u}$, $\tilde{d}$, $\tilde{s}$, $\tilde{c}$, $\tilde{b}$)
with  nearly mass-degenerate left and right helicity states.
Production of top squarks is not considered even though
 the lighter of the two top-squark mass eigenstates 
can be lighter than the other squarks \cite{light_stop}.
We impose common scalar and gaugino masses 
at a GUT scale as in the minimal supergravity model~\cite{sugra},
and use the renormalization group equations~\cite{RGE} 
to generate the slepton masses 
which require $\msquark\ \gtsim\ 0.9 \mgluino$. 
To avoid a region in MSSM parameter space where there are significant
 branching ratios of chargino and neutralino decays into Higgs particles,
 the pseudoscalar Higgs mass is set to 500 \gevcc\ which is 
above the \chione\    and \chitwo\  masses.
With these assumptions, the sensitivity of our
search can be studied as a function of four parameters: 
the gluino mass (\mgluino),
the squark mass (\msquark), 
the ratio of the vacuum expectation values
of the two Higgs fields ($\tan\beta$), and 
the Higgs mass  parameter ($\mu$). 
Since we choose to decouple our search from the
Higgs sector we scan a  range of $\mu$ that is both consistent with
LEP results~\cite{ALEPH,LEP_mu} and  less than 
the SUSY mass scale: $ 100\ \ltsim\ | \mu | ~\ltsim\ 1000\ \gevcc$.

The acceptance for SUSY processes is estimated by performing the final data 
selection on events simulated  with \ISAJET~\cite{isajet} 
using CTEQ3L~\cite{CTEQ} parton distribution functions (PDFs). 
These events are then passed  through the CDF detector simulation.
We define the acceptance as the ratio of the number of 
dilepton events that pass our cuts to the total number of generated SUSY events
which contain at least two leptons. 
For a nominal SUSY scenario of  \mgluino\ = 200~\gevcc,
infinite squark mass (and hence infinite slepton mass), $\tan \beta = 2$ and 
$\mu=-800$~\gevcc, the  acceptance is $1\%$, 
due mostly to the lower  \pt\ values of the leptons. For the case where 
$\msquark \simeq \mgluino \simeq 200$~\gevcc, 
the slepton (\sleptonR)  mass is  lighter. 
This  enhances 
the leptonic branching ratio due to $\chitwo \to \ell \sleptonR$,
resulting in an increase of LS dilepton events
in $\gluino\gluino,\gluino\squark, \squark\squarkb$ production
and 
a rise of the overall acceptance to $3\%$.
Table~\ref{tab:datared} and
Figure~\ref{fig:data} compare the data reduction, 
the expectations from SM processes, and a SUSY scenario.

The total systematic uncertainty on the expected number of 
LS signal events comes from uncertainties on the theoretical calculation
of the production cross section of gluinos and squarks, the event acceptance,
and the integrated luminosity. 
The NLO cross section depends mainly
on the choices of the QCD renormalization scale ($Q^2$) and
PDFs~\cite{PROSPINO}.
The nominal choice of $Q^2$ is $m^2$, where $m$ is
\msquark,  \mgluino, or
$\frac{1}{2}\sqrt{\msquark^{2} + \mgluino^{2}}$ for
$\squark\squark$/$\squark\bar{\squark}$, $\gluino\gluino$, or
$\squark\gluino$ production, respectively. 
The uncertainty due to
the choice of $Q^2$ is determined to be 21\%  by taking
the larger of the variation of the cross section 
at  $Q^2 = (m/2)^2$ and at $Q^2 = (2m)^2$
from the nominal cross section value.
Similarly, the variation of the cross section due to the choice of PDFs 
yields an 8\% uncertainty, estimated as the maximum deviation
between the nominal choice of CTEQ3M~\cite{CTEQ} and 
MRS(G)~\cite{MRSG} or GRV94HO~\cite{GRV}.
Uncertainty on the signal
acceptance is due to uncertainties on the efficiencies
of the lepton trigger, identification and
isolation efficiencies, as well as on the jet energy scale and the
amount of gluon radiation.  
By varying the measured lepton trigger
and identification efficiencies by one standard deviation,
the acceptance uncertainties are estimated to be 5\% and 3\%, respectively.  
Since the lepton isolation efficiency depends on jet multiplicity, 
the uncertainty  is estimated
using $Z^0(\to \ell^+ \ell^-)\ + \ge 2$~jet events and is found to be  11\%. 
By varying the jet energy scale by one standard deviation, 
we find a 5\% effect on the acceptance.
The uncertainty due to the initial and final state gluon radiation
(ISR and FSR) is estimated by turning the ISR and/or FSR radiation off,
which gives at most 7\% variation in the acceptance.
Enough MC events are generated so as to keep
the statistical uncertainty below 3\%. 
The uncertainty on the luminosity is 4\%.
The  combined uncertainty is calculated by adding all uncertainties
in quadrature, and  is found to be 28\%.

\begin{figure}
\begin{center}
\vspace*{-0.5cm}
\epsfig{file=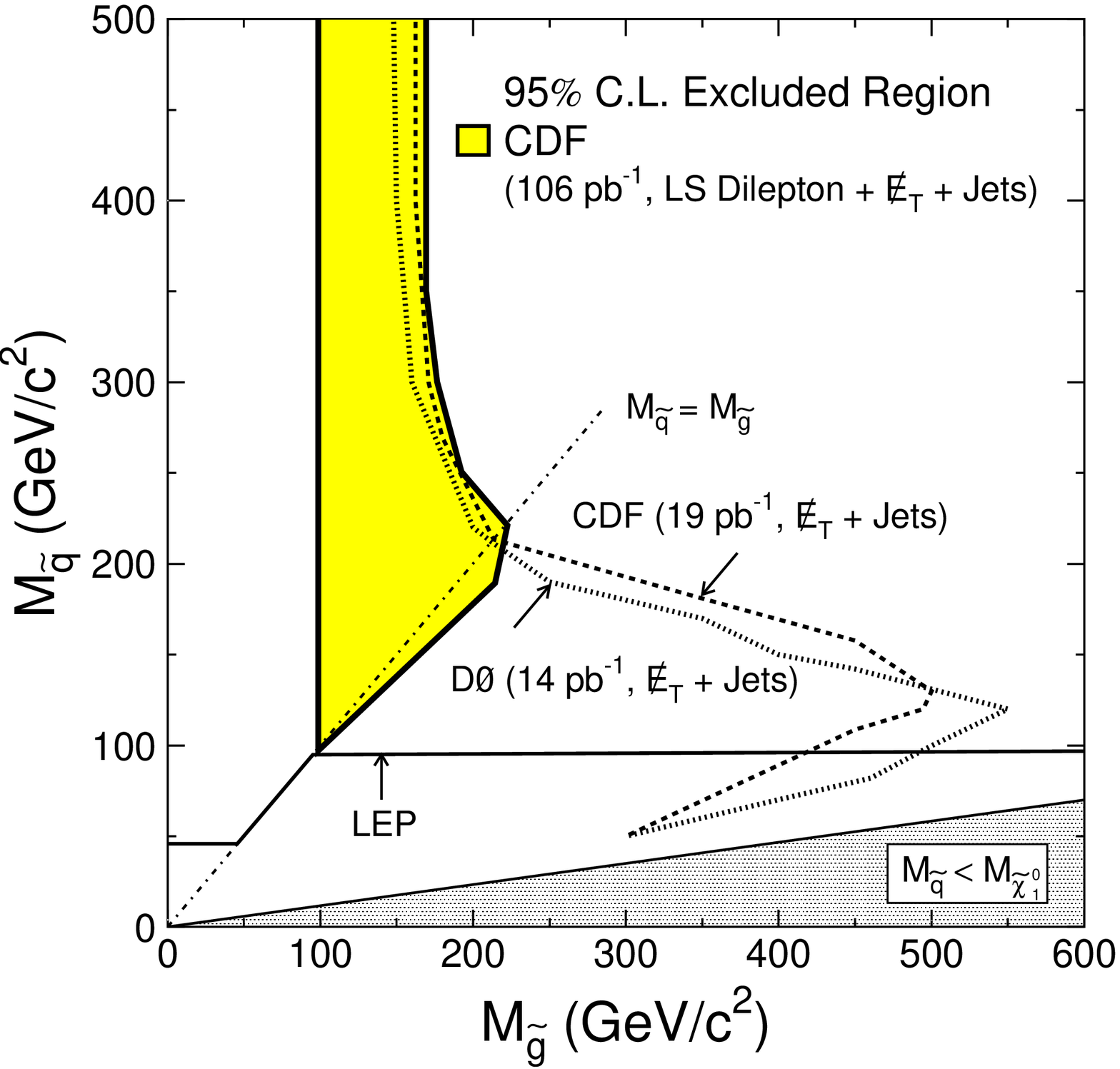,width=220pt} 
\caption{Limit in the \msquark-\mgluino\ plane
at the 95\% confidence level for 
a constrained MSSM scenario (\msquark\ $\protect\gtsim$ 0.9 \mgluino)
for $\tan \beta = 2$ and $\mu = -800$~\gevcc.  
The results of other direct, but complementary,  searches are
also presented
\protect\cite{L3,ALEPH,CDF_MET_0_IA,D0_MET_IA}.}	
\vspace*{+0.2cm}
\label{fig:limit_contour}
\end{center}
\end{figure}

Since no LS events pass our cuts, 
we calculate the upper limit on the number of SUSY events
at  the 95\% confidence level (C.L.)
using a frequentist algorithm~\cite{limits}
with a systematic uncertainty of 28\%
and no background subtraction.
This corresponds to 3.46 events which we use 
to exclude regions  in the \msquark-\mgluino\ plane.
Figure~\ref{fig:limit_contour} shows the exclusion region 
for $\tan \beta = 2$ and $\mu = -800$~\gevcc. 
We set 95\% C.L. limits at $\mgluino > 168\ \gevcc$
for $\msquark \gg \mgluino$ and
$\mgluino > 221\ \gevcc$ for $\mgluino \simeq \msquark$.
These results are better than the previous
limits from complementary searches by about 5 \gevcc\ 
\cite{CDF_MET_0_IA,D0_MET_IA}.

We examine the dependence of the mass limit as $\tan \beta$ and $\mu$ are
varied in the region $\mgluino \simeq \msquark$.
For $\mu = -800\ \gevcc$,
the variation in the mass limit
is smaller than 2\% in the range of $\tan \beta$ between
1.7 and 10 if the mixings of the third generation SUSY particles
(especially $\stau$) are minimal.
In the case of maximal $\tilde{\tau}$-mixing, 
the mass limit remains the same for $\tan \beta$ up to about  3. 
For $\tan\beta = 2$,
the limit deviates by at most 3.6\% from the 221 \gevcc\ limit
in the range $\mu \le -150\ \gevcc$,
while
the limits in $\mu \ge 150$~\gevcc\ are
systematically 8-12\% lower.

In conclusion, we have searched for new physics using
LS dilepton events in association with two or more jets and $\met$ 
in \ppbar\ collisions at $\sqrt{s}$ = 1.8 TeV. 
Production of both OS and LS dilepton events are
consistent with the SM expectations.
Within a framework of constrained MSSM (5 degenerate squarks,
\msquark\ $\gtsim$ 0.9 \mgluino), 
for small $\tan\beta$ we set mass limits of
$\mgluino > 168$ \gevcc\ for  $\msquark \gg \mgluino$,
and $\mgluino > 221$~\gevcc\ for  $\msquark \simeq \mgluino$,
both with small variation as a function of $\mu$.

%

We thank the Fermilab staff and the technical staffs 
of the participating institutions for their vital contributions.  
This work was supported by the U.S. Department of Energy and
National Science Foundation;
the Italian Istituto Nazionale di Fisica Nucleare;
the Ministry of Education, Science, Sports and Culture of Japan; 
the Natural Sciences and Engineering Research Council of Canada; 
the National Science Council of the Republic of China; 
the Swiss National Science Foundation; 
the A.~P.~Sloan Foundation;  
the Bundesministerium f\"{u}r Bildung und Forschung, Germany; 
the Korea Science  and Engineering Foundation (KoSEF); 
the Korea Research Foundation;
and the Comision Interministerial de Ciencia y Tecnologia, Spain.

\def\Journal#1#2#3#4{{#1} {\bf #2}, #3 (#4)}
\def\NCA{Nuovo Cimento}
\def\NIM{Nucl. Instrum. Methods}
\def\NIMA{{Nucl. Instrum. Methods} A}
\def\NPB{{Nucl. Phys.} B}
\def\PLB{{Phys. Lett.}  B}
\def\PRL{Phys. Rev. Lett.}
\def\RPP{Rep. Prog. Phys.}
\def\PRD{{Phys. Rev.} D}
\def\PR{Phys. Rep.}
\def\ZPC{{Z. Phys.} C}
\def\MPL{{Mod. Phys. Lett.} A}
\def\EPJC{{Eur. Phys. J.} C}
\renewcommand{\baselinestretch}{1}

\begin{table}
\caption{A comparison of the event reduction for the data,
Standard Model (SM) backgrounds and a model of SUSY production with 
$\tan \beta = 2$, $\mu = -800$~\gevcc, $\mgluino = 210$~\gevcc, and $\msquark = 211$~\gevcc.}
\begin{center}
\begin{tabular}{l r r r}
\multicolumn{1}{c}{Selection}  &
\multicolumn{1}{c}{Data}   &
\multicolumn{1}{c}{SM Backgounds}     &
\multicolumn{1}{c}{SUSY}   \\
\hline
Dilepton Dataset                   & 515,699 & & \\
Dilepton-Dijet                     &  350 & & \\
$M_{\ell\ell} \ge 12$ \gevcc~       &  255 & $279\pm9\pm79$ & $27\pm1\pm5$\\
$Z^0 (\rightarrow \ell^+ \ell^-)$ veto 
                                   & 128 & $158\pm7\pm45$ & $27\pm1\pm5$\\
$\met \ge 25$ GeV              & 19 & $14.7\pm1.3\pm2.8$ & $24\pm1\pm5$\\
Like-sign Dilepton          & 0 & $0.55\pm0.25\pm0.08$ & $5.9\pm0.6\pm1.4$\\
\end{tabular}
\end{center}
\label{tab:datared}
\end{table}

\begin{table}
\caption{The expected backgrounds from Standard Model
contributions to the final data selection after all
but the LS requirement in Table \ref{tab:datared}.  
Opposite-sign and like-sign  dilepton events are listed.}
\begin{center}
\begin{tabular}{l r r }
\multicolumn{1}{c}{Source}     & \multicolumn{1}{c}{Opposite-sign}
&
\multicolumn{1}{c}{Like-sign}  \\
\hline
Drell-Yan           & ~8.7 $\pm$ 0.9 $\pm$ 2.5 &
\multicolumn{1}{l}{0.00~~$^{+0.01}_{-0.00}$~~$^{+0.01}_{-0.00}$}\\
$t\bar{t}$          & ~4.0 $\pm$ 0.3 $\pm$ 1.2 & 0.08 $\pm$ 0.04 $\pm$ 0.02\\
$b\bar{b}/c\bar{c}$ & ~0.9 $\pm$ 0.9 $\pm$ 0.3 & 0.23 $\pm$ 0.23 $\pm$ 0.07\\
Diboson
& ~0.5 $\pm$ 0.1 $\pm$ 0.1 & 0.24 $\pm$ 0.10 $\pm$ 0.04\\
\hline
Total               & 14.1 $\pm$ 1.3 $\pm$ 2.8 & 0.55 $\pm$ 0.25 $\pm$ 0.08\\
Data                &                   19     &       0 \\
\end{tabular}
\end{center}
\label{tab:SM_BG}
\end{table}

\newpage

\begin{thebibliography}{99}
\bibitem{cdfdefs}
	We use a coordinate system where $\theta$ and $\phi$ are the polar
  and azimuthal angles, respectively, with respect to the proton
  beam direction ($z$ axis). The pseudorapidity $\eta$ is defined as
  $-\ln [ \tan(\theta/2) ]$.  
  The transverse momentum of a particle is denoted as 
  $\pt = p \sin\theta$.  
  The missing transverse energy, $\met$, is the magnitude of
  $\metvec \equiv -\sum E_T^i \,\hat{n}_i$, 
  where $\hat{n}_i$ is the unit vector in the
  transverse plane pointing from the interaction point to
  the energy deposition in calorimeter cell $i$.
\end{thebibliography}
\end{document}